# Liquid-liver phantom: mimicking the viscoelastic dispersion of human liver for elastography in ultrasound and MRI

Anna S. Morr[1], Helge Herthum[1,2], Felix Schrank[1], Steffen Görner[1], Matthias S. Anders[1], Markus Lerchbaumer[1], Hans P. Müller[1], Thomas Fischer[1], Klaus-Vitold Jenderka[3], Hendrik H.G. Hansen[4], Paul A. Janmey[5], Jürgen Braun[2], Ingolf Sack[1], Heiko Tzschätzsch[1]

1. Department of Radiology, Charité – Universitätsmedizin Berlin, Corporate Member of Freie Universität Berlin and Humboldt - Universität zu Berlin, Charitéplatz 1, 10117, Berlin, Germany.
2. Institute of Medical Informatics, Charité – Universitätsmedizin Berlin, Corporate Member of Freie Universität Berlin and Humboldt - Universität zu Berlin, Charitéplatz 1, 10117, Berlin, Germany.
3. Department of Engineering and Natural Sciences, University of Applied Sciences Merseburg, Eberhard-Leibnitz-Str. 2, 06217, Merseburg, Germany.
4. Medical Ultrasound Imaging Center, Department of Medical Imaging, Radboud University Medical Center, Nijmegen, The Netherlands.
5. Institute for Medicine and Engineering, Center for Engineering Mechanobiology, and Department of Physiology, University of Pennsylvania, Philadelphia, PA, USA.

E-mail: Ingolf.sack@charite.de

**Abstract**

Different clinical elastography devices show different liver-stiffness values in the same subject, hindering comparison of values and establishment of system-independent thresholds for disease detection. Therefore, authorities request standardized phantoms that address the viscosity-related dispersion of stiffness over frequency.

A linear polymerized polyacrylamide phantom (PAAm) was calibrated to the viscoelastic properties of healthy human liver *in vivo*. Shear-wave speed as a surrogate of stiffness was quantified between 5 Hz and 3000 Hz frequency-range by shear rheometry, ultrasound-based time-harmonic elastography, clinical MR elastography (MRE), and tabletop MRE. Imaging parameters for ultrasound were close to those of liver *in vivo*. Reproducibility, aging behavior and temperature dependency were assessed and fulfilled requirements for quantitative elastography. In addition, the phantom was used to characterize the frequency bandwidth of shear-wave speed of several clinical elastography methods.

The liquid-liver phantom has favorable properties for standardization and development of liver elastography: first, it can be used across clinical and experimental elastography devices in ultrasound and MRI. Second, being a liquid, it can easily be adapted in size and shape to specific technical requirements, and by adding inclusions and scatterers. Finally, since the phantom is based on non-crosslinked linear PAA constituents, it is easy to produce, indicating potential widespread use among researchers and vendors to standardize liver-stiffness measurements.

**Abbreviations**

LS – Liver stiffness

MCC – Microcrystalline cellulose

MRE – Magnetic-resonance elastography

MRI – Magnetic-resonance imaging

PAAm – Polyacrylamide

QIBA – Quantitative Imaging Biomarkers Alliance

SWS – Shear-wave speed

THE – Time-harmonic elastography

USE – Ultrasound elastography

**Introduction**

Liver stiffness (LS) is an established imaging marker for hepatic fibrosis. LS is an intrinsic material property and can be quantified *in vivo* by elastography and *ex vivo* by standardized reference methods such as shear rheometry. Nevertheless, there are no generally accepted system-independent LS thresholds for the detection of liver fibrosis. Today, different clinical elastography devices show different values for shear-wave speed (SWS), a surrogate of LS, in the same volunteer[1]. This has been identified by experts, including the Quantitative Imaging Biomarkers Alliance (QIBA), as a major obstacle to the standardization and establishment of quantitative LS markers. Recently, QIBA compared different clinical elastography methods in standardized phantoms and concluded that all methods could distinguish correctly between soft and firm phantoms[2], but inter-device deviations increased in presence of viscosity such as that of human liver *in vivo*[1,3]. Variability of LS due to viscosity is introduced by SWS dispersion, i.e. increase of SWS over excitation frequency. Consequently, SWS can only be compared between modalities when the excitation frequency is precisely known or when measurements are performed by the same excitation method. However, most methods are transient, resulting in travelling wave packages that have a group frequency corresponding to a wider spectrum of shear-wave frequencies. Even more challenging, this group frequency decreases with travel time owing to viscous damping. Hence, it is important to characterize the group frequency of an elastography device by means of a phantom with realistic and precisely known SWS dispersion properties. Such a liver-mimicking SWS-dispersion phantom should be visible in ultrasound and MRI, in order to be accessible to both ultrasound elastography (USE) and magnetic-resonance elastography (MRE). Current commercial phantoms do not meet these criteria as they are purely elastic[4] and not equally well accessible by USE and MRE.

We here introduce an SWS-dispersion phantom made of linear polymerized polyacrylamide (PAAm), which is visible by USE and MRE. In the past, PAAm gels have been used as matrices with tunable viscoelastic properties for biomedical research[5,6]. The dispersion of SWS and shear-wave damping (supplementary material) of our phantom was tuned to that of healthy human liver, as reported in literature over a wide frequency range[7-13] and characterized by shear rheometry and experimental multifrequency elastography. The aim of this study was to foster the standardization of clinical elastography across devices and platforms by using a comprehensively characterized phantom that mimics SWS-dispersion of the human liver and can be generated reproducibly.

**Methods**

*Phantoms*

PAAm phantoms of 2L total volume were made by dissolving 400g acrylamide in 1565ml distilled water. Polymerization was imitated by adding 34g ammonium persulfate and 45ml tetramethylethylenediamine. After stirring, the phantoms were placed in a cubic box with a spaghetti vinyl mat (Mattenlager, Germany) at the bottom to minimize potential acoustic reflection (Figure 1). All phantoms were stored for two months in the dark at 21°C for curing, to ensure stability. To provide ultrasound speckle contrast, 5g microcrystalline cellulose (0.5μm, MCC, Acros Organics, USA) was added. Three identical phantoms were produced on three different days. Imaging parameters and SWS were compared with those of the CIRS phantom (Elasticity QA phantom model 049, CIRS, Norfolk, USA).

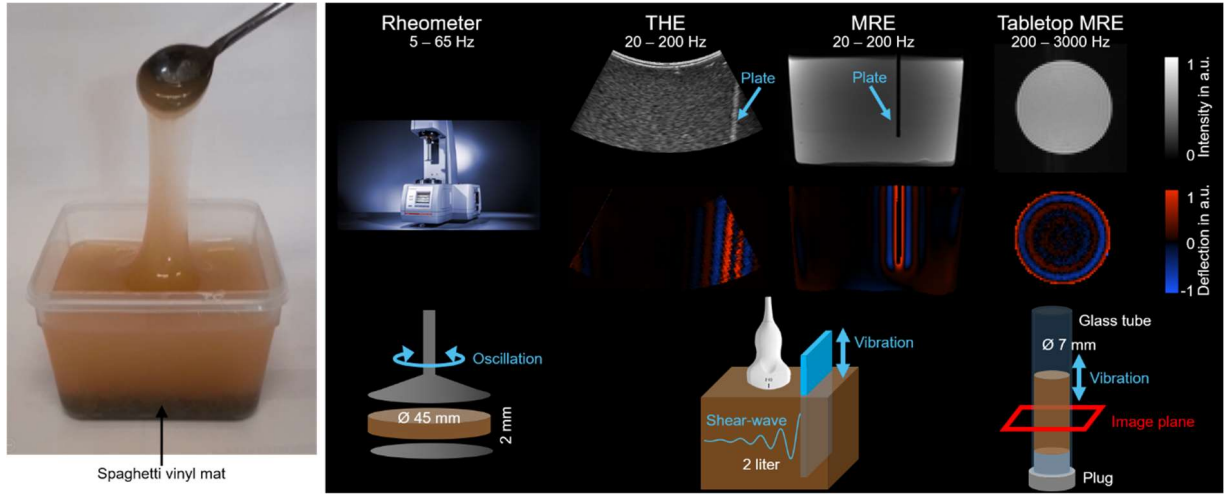

*Figure 1* Left: liquid-liver phantom based on linear polymerized polyacrylamide (PAAm) with demonstration of its viscous property. Right: overview of the modalities used for characterization of the viscoelastic properties of the PAAm phantom from 5 to 3000 Hz. Top row: standard image contrast, middle row: shear-wave deflection fields, bottom row: experimental setup. Time-harmonic elastography (THE), magnetic-resonance elastography (MRE). The image of the rheometer device was kindly made available by Anton Paar GmbH, Graz, Austria.

*Reference elastography*

SWS dispersion and shear-wave damping (supplementary material) in PAAm were quantified from 5 to 65Hz at 21°C by shear oscillatory rheometry (Modular Compact Rheometer 301, Anton Paar, Austria), from 20 to 200Hz at 21–22°C by multifrequency MRE (3-Tesla Magnetom Lumina, Siemens, Erlangen, Germany) and ultrasound-based time-harmonic elastography (THE, SonixMD, Ultrasonix, Scottsdale, USA) and from 200 to 3000Hz at 26°C by tabletop MRE (0.5-Tesla MagSpec, Pure Devices, Würzburg, Germany)[14]. In THE and MRE, plane shear waves were introduced by a vibration plate immersed in the phantom while tabletop MRE induced cylindrical waves in a glass tube. The experimental setups are illustrated in Figure 1.

*Clinical elastography*

The following clinical USE and MRE modalities were used: shear wave elastography (Aplio i900, Canon, Tokyo, Japan) at 3.5cm depth, Virtual Touch™ (Acuson Sequoia, Siemens, Pennsylvania, USA) at 2.7cm depth and FibroScan® (Echosense, Paris, France), liver THE (six superimposed frequencies, 27-56Hz, mean 41.5Hz)[13] and liver MRE[12] (four consecutive frequencies, 30-60Hz, mean 45Hz). For FibroScan®, a thin silicon mat was stretched tightly over the phantom to provide a counterforce to the transducer. For THE, the phantom was placed on a vibration bed[13] and scanned from the top, while for MRE the phantom was placed in a head coil on two air-bottle drivers[15]. Further details of imaging parameters are given in the supplementary material.

*Data-processing*

Magnitude $|G^*|$ and phase angle of the complex shear modulus $\phi = \arg(G^*)$ obtained from shear rheometry were converted to SWS by

$$SWS = \sqrt{\frac{2\,|G^*|}{\rho\,[1+\cos(\phi)]}}, \qquad (1)$$

with PAAm density $\rho$=(1.094±0.006)g/cm³. Plane shear waves of frequency 20–200Hz acquired by THE and MRE were analyzed by fitting a 2-dimensional shear-wave with exponential decay to the complex wave at drive frequency. A Bessel function fit was used for analyzing 200–3000Hz cylindrical waves from tabletop MRE[14]. SWS dispersion was analyzed by fitting to the data the two-parameter springpot model,

$$G^* = \mu^{1-\alpha} \, (i\, \omega\, \eta)^{\alpha}, \qquad (2)$$

assuming a viscosity $\eta$=1Pa·s. $\omega$ denotes angular drive frequency, while $\mu$ and $\alpha$ denote shear modulus and viscoelastic powerlaw coefficient, respectively. $G^*$ was converted to SWS according to eq.(1). Clinical elastography data were processed by standard methods (adjusted filter for THE[13] and MRE[12]) on the different platforms and converted to SWS accordingly.

*Reproducibility, aging, temperature effect and effect of MCC*

In three different phantoms, SWS was measured 2 months after phantom production across the full range of frequencies, by all modalities, to check the reproducibility of phantom preparation. The effects of aging, temperature and MCC on SWS were assessed by tabletop MRE in the frequency range 200–3000Hz. To assess aging, SWS was measured in one phantom without MCC over 17 days, starting directly after production. Difference-SWS was measured in three phantoms, between 26 and 37°C and with and without MCC for each frequency and averaged to direct effect sizes and corresponding effect at 30Hz using eq.(2).

*Statistics*

Statistics were performed with GraphPad Prism 9.0 (GraphPad software, USA). Statistical significance was defined by $p<0.05$. For temperature and MCC effect, a two-tailed paired *t*-test was performed. For aging, repeated-measure one-way ANOVA with *post-hoc* Tukey's multiple comparisons test was used. The rheological model was fitted with MATLAB R2019b (MathWorks Inc., USA), and standard error was calculated by bootstrapping. Values are presented either as mean±standard error or as median [range].

**Results**

*Phantom characterization*

Mean SWS of all PAAm phantoms acquired by rheometry, MRE, and THE along with a springpot fit according to eq.(2) are shown in Figure 2. The fit with $\mu$=(6.769±0.023)kPa and $\alpha$=0.3671±0.0023 is redrawn in Figure 3 superposed over literature data for healthy human liver and commercial CIRS and Resoundant (MRE Phantom, Rochester, USA) phantoms.

At reference frequency of 30Hz, which lies in the middle of group frequencies of all elastography methods ((17–45)Hz, see below), the springpot predicts an $SWS_{30Hz}$ of (1.345±0.005)m/s. $SWS_{30Hz}$ reproducibility in three different phantoms was (1.30±0.03)%. Within the first seven days after phantom production, $SWS_{30Hz}$ increased significantly by (5.75±0.20)%, while between days 7 and 17, $SWS_{30Hz}$ changed only insignificantly by (0.71±0.16)% (Fig.4). In the temperature range (26–37)°C, $SWS_{30Hz}$ decreased significantly by (1.136±0.005)%/°C. Adding MCC significantly increased $SWS_{30Hz}$ by (1.12±0.04)%. All parameters are summarized in Table 1 (see supplementary material for the corresponding analysis of shear-wave damping).

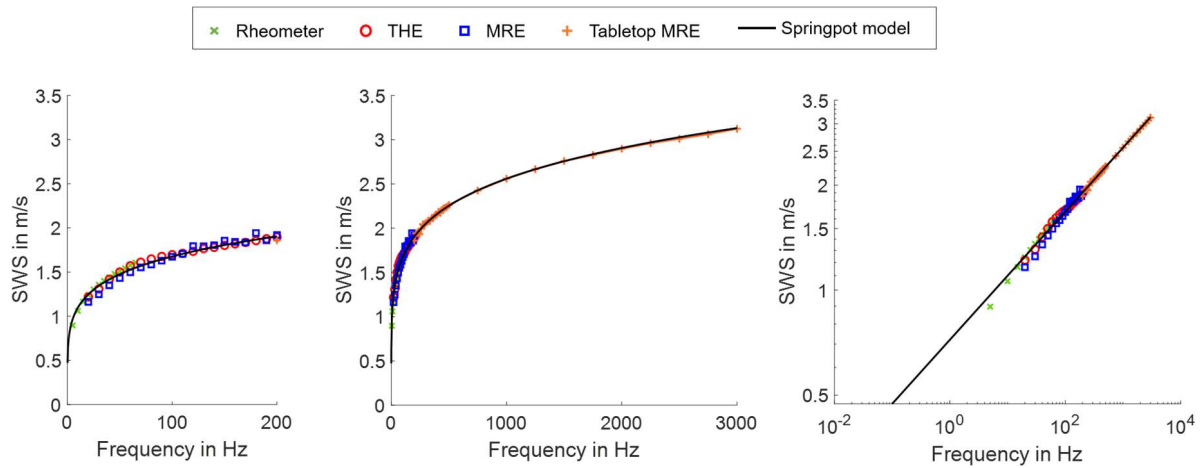

*Figure 2* Multimodal characterization of shear-wave speed (SWS) dispersion of the polyacrylamide (PAAm) phantom with springpot fit (black line). Left-hand side: frequency zoom 0 to 200 Hz; middle: full frequency range, normal scale; right-hand side: full frequency range, log scale. Time-harmonic elastography (THE), magnetic-resonance elastography (MRE)

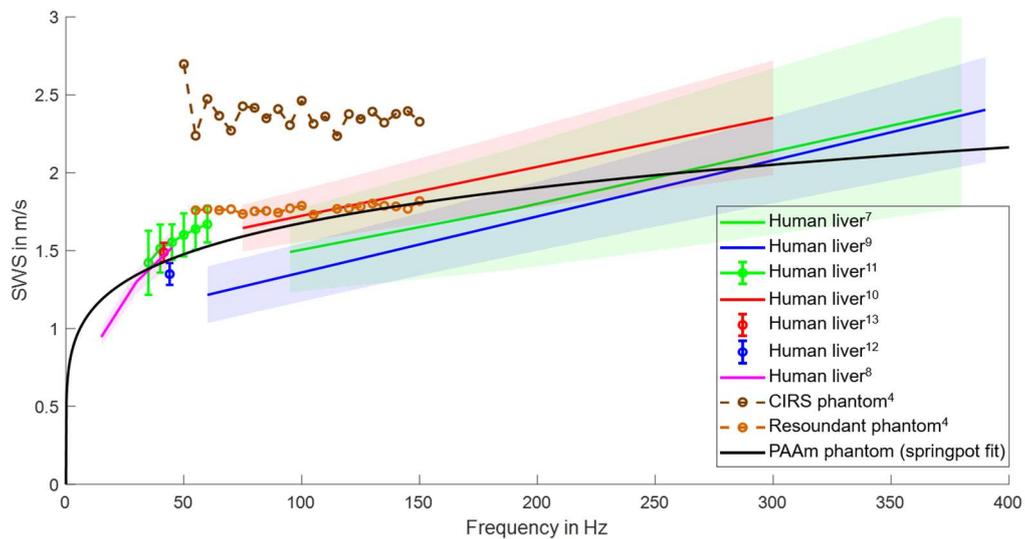

*Figure 3* Ranges of literature data of frequency-resolved shear-wave speed (SWS) with corresponding standard deviation of healthy human liver in vivo and of commercial elastography phantoms. In addition, the fit of SWS dispersion of the proposed polyacrylamide (PAAm) phantom as shown in figure 2 is re-plotted (black line).

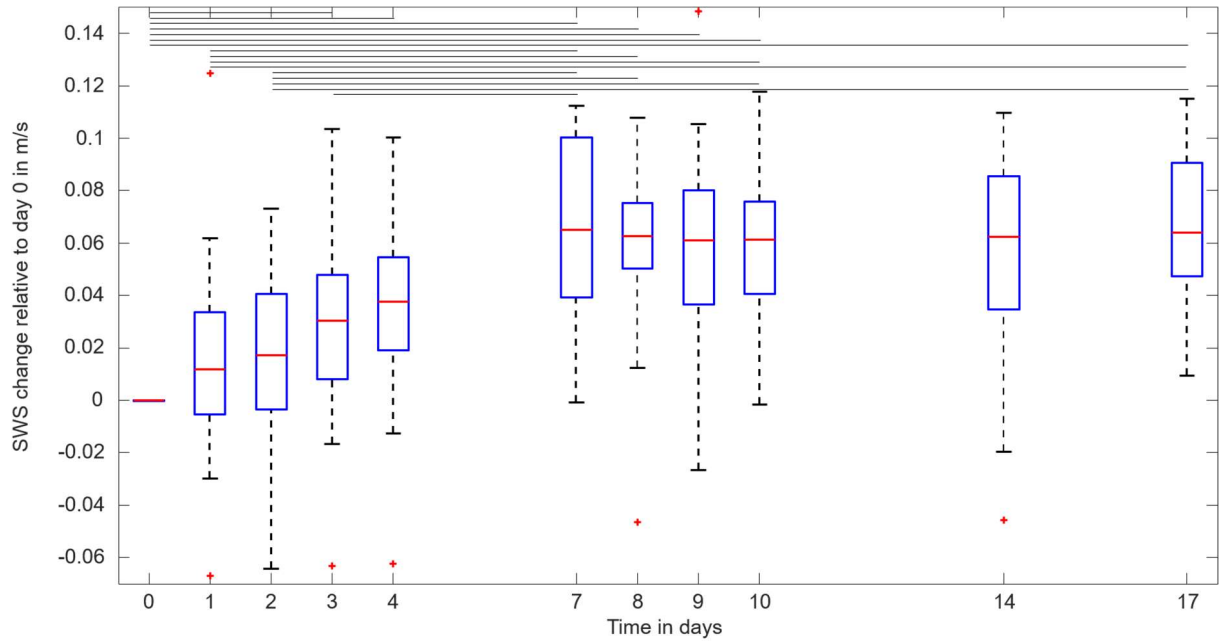

*Figure 4* Analysis of shear-wave speed (SWS) changes due to aging based on tabletop MRE in the frequency range from 200 to 3000Hz. SWS changes relative to production (day 0) are shown. The variability of data due to frequencies is indicated by the boxplots. Horizontal black lines demarcate significant changes.

| Group | Type | Parameter | Unit | Mean | SE | P-value | SWS at 30 Hz | Requirements |
|---|---|---|---|---|---|---|---|---|
| Viscoelastic properties | Rheological model | $\alpha$ | | 0.3671 | 0.0023 | | 1.345 ± 0.005 m/s | |
| | | $\mu$ | kPa | 6.769 | 0.023 | | | |
| | | | | | | | **Relative effect at 30 Hz** | |
| | Reproducibility | SWS range | m/s | 0.0349 | 0.0004 | | 1.30 ± 0.03 % | ± 5 % |
| | Aging effect | $SWS_{Day7} - SWS_{Day0}$ | m/s | 0.077 | 0.003 | 0.0004 | 5.75 ± 0.20 % / 7 days | < 5 % / 6 month |
| | | $SWS_{Day17} - SWS_{Day7}$ | m/s | 0.0095 | 0.0021 | 0.999 | 0.71 ± 0.16 % / 10 days | |
| | Temperature effect | $SWS_{37°C} - SWS_{26°C}$ | m/s/°C | -0.01528 | 0.00007 | < 0.0001 | - 1.136 ± 0.005 % / °C | 21 ± 1 °C |
| | MCC effect | $SWS_{MCC} - SWS_{No\,MCC}$ | m/s | 0.0151 | 0.0006 | < 0.0001 | 1.12 ± 0.04 % | ± 5 % |
| | | | | **Median** | **Range** | | **Assigned frequency** | |
| Imaging properties | Ultrasound | Attenuation | dB/cm/MHz | 0.38 | [0.37 - 0.43] | | | 0.5 ± 0.1 dB/cm/MHz |
| | | Speed of sound | m/s | 1646 | [1627 - 1656] | | | 1540 ± 30 m/s |
| | MRI | T1 | ms | 1434 | [1428 - 1442] | | | 812 ± 65 ms* |
| | | T2 | ms | 213 | [212 - 215] | | - | 42 ± 3 ms* |
| Clinical elastography | Aplio | SWS | m/s | 1.24 | [1.11 - 1.25] | | 20 [11 - 20] Hz | |
| | Acuson | SWS | m/s | 1.37 | [1.37 - 1.44] | | 33 [33 - 43] Hz | |
| | FibroScan | SWS | m/s | 1.21 | [1.19 - 1.24] | | 17 [16 - 19] Hz | |
| | THE | SWS | m/s | 1.41 | [1.39 - 1.43] | | 41.5 Hz | |
| | MRE | SWS | m/s | 1.45 | [1.45 - 1.46] | | 45 Hz | |

*Table 1* Experiments, modalities and effects addressed in this study of shear wave speed (SWS) dispersion of the liquid-liver phantom. Additionally, specifications from the Quantitative Imaging Biomarkers Alliance17 and values from healthy human liver in vivo20 (*) are given. Time-harmonic elastography (THE), Magnetic-resonance elastography (MRE), Microcrystalline cellulose (MCC) for ultrasound contrast, Standard error (SE).

*Clinical elastography*

SWS from clinical elastography is shown in Figure 5. Values are 1.24 [1.11–1.25]m/s for Aplio, 1.37 [1.37–1.44]m/s for Acuson, 1.21 [1.19–1.24]m/s for FibroScan, 1.41 [1.39–1.43]m/s for THE and 1.45 [1.45–1.46]m/s for MRE (Table 1). On the basis of the springpot fit shown in Figure 2, group frequencies for transient excitation methods were characterized as 20 [11–20]Hz (Aplio), 33 [33–43]Hz (Acuson) and 17 [16–19]Hz (FibroScan). SWS values obtained by harmonic excitation methods were equal to those predicted by the fit (Fig.6).

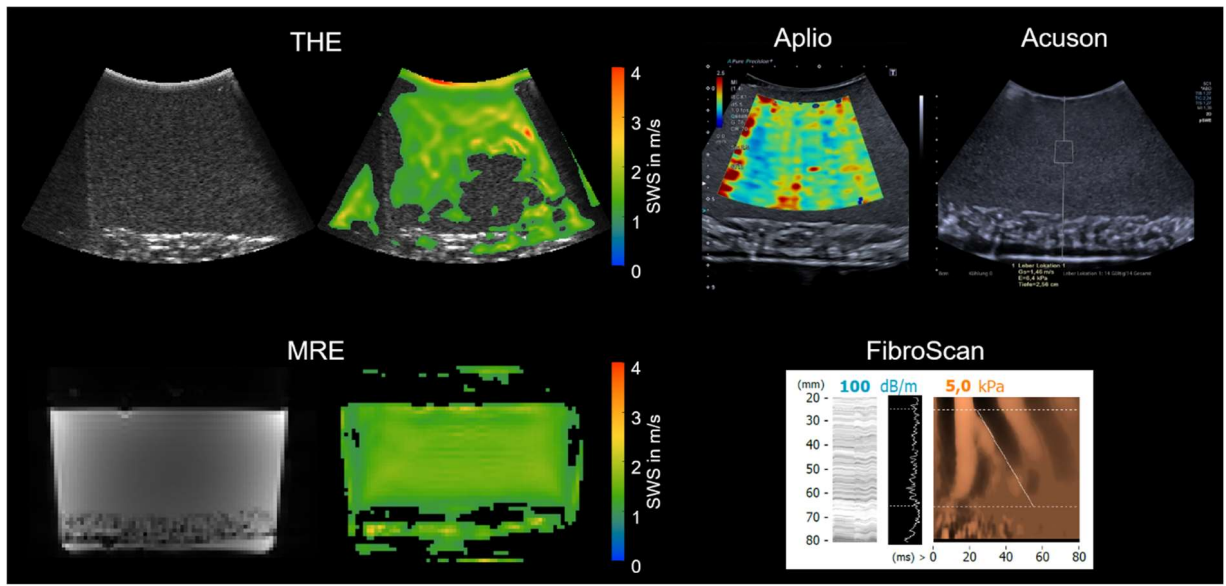

*Figure 5* Clinical elastography. Representative ultrasound images and MRI with shear-wave speed (SWS) maps obtained by Time-harmonic elastography (THE), magnetic-resonance elastography (MRE), Aplio from Canon, Acuson from Siemens and FibroScan from Echosense.

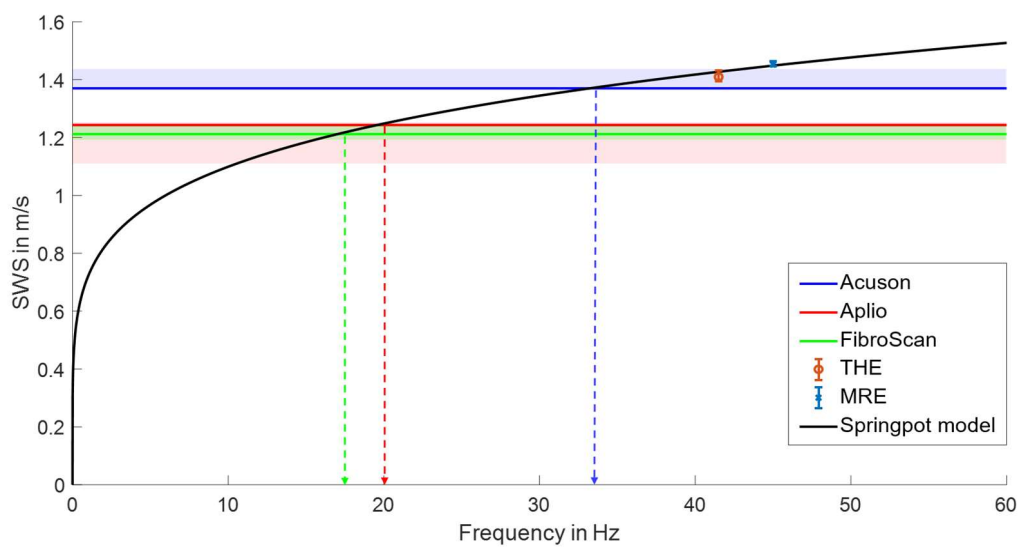

*Figure 6* Assignment of the effective frequency (dashed lines) of clinical elastography by the liquid-liver phantom. Median shear-wave speed (SWS) values and ranges obtained by clinical elastography with transient shear wave excitation (Acuson, Aplio, and FibroScan) are shown as horizontal lines. Median values and ranges of clinical elastography based on harmonic shear wave excitation (THE and MRE) are shown by markers at mean harmonic frequency. Time-harmonic elastography (THE), magnetic-resonance elastography (MRE)

*Imaging properties*

Attenuation of ultrasound waves was 0.38 [0.37–0.43]dB/cm/MHz while speed of sound was 1646 [1627–1656]m/s. Visual speckle brightness was higher compared with the CIRS phantom (supplementary Figure 1). T2-weighted MRI was more homogenous in the PAAm phantom than in the CIRS phantom. T1 and T2 relaxation times were 1434 [1428–1442]ms and 213 [212–215]ms (Table 1).

**Discussion**

The PAAm phantom introduced here is the first phantom for USE and MRE which mimics the SWS dispersion of healthy human liver[7-13]. Viscoelastic properties were comprehensively characterized over a wide frequency range of nearly 3000Hz. We demonstrated the use of our phantom for measurement of effective group frequency in devices with transient excitation. This investigation provided important information towards comparing stiffness values among different systems.

Although our PAAm phantom well reproduces the viscoelastic dispersion of healthy human liver (above 30Hz), it has no static shear modulus. This prompts the term *liquid-liver phantom* although, clearly, the liver is not a liquid. Nevertheless, the fact that the phantom behaves as a liquid at static load, while fitting literature data for solid-liver properties above 15Hz, indicates a superviscous material behavior similar to that recently reported for human brain[16].

It is a salient finding of our study that the phantom can be produced reproducibly, with 1.3% variability, and thus fulfills the QIBA requirements of 5%[17]. Given recent reports about the excellent long-term stability of crosslinked PAAm[18] combined with our observation of unchanged properties from week 1-2 of curing, we are confident that our phantom changes less than 5% over 6 month as requested by QIBA[17]. Also, the change of SWS with temperature was low, with only 1.1% for each °C within our test range.

Compared with other elastography phantoms from the literature, the liquid-liver phantom uniquely combines realistic viscoelastic dispersion properties, multi-modality and ease of reproducible production. Current commercial phantoms are purely elastic[4] and are either limited in MRI (CIRS) or USE (Resoundant). Chatelin *et al.* have proposed using polyvinyl chloride gels for phantoms in USE and MRE, but without reporting viscous properties[19]. A viscoelastic phantom used by QIBA[3] gave an SWS dispersion similar to that of human liver, though without information on shear-wave damping and with limited reproducibility of the production.

*Limitations*

Although SWS dispersion observed in our phantom agreed well with that of human liver, the shear-wave damping was more pronounced than in the liver. To better mimic shear-wave damping of the human liver, PAAm could be produced based on a mixture of crosslinked and linearly polarized chains, as demonstrated preliminarily (see supplementary material). However, it remains to be determined whether such partially crosslinked PAAm phantoms can be reproducibly fabricated and provide the same flexibility for adding wave sources and scatterers as demonstrated herein. It should be noted that MCC particles segregate and sink under gravity within weeks. We solved this problem either by stirring before investigations or by placing the phantom in a plastic bag and kneading it directly before measurement, to redistribute the scatterers uniformly (see supplementary material). Finally, the speed of sound and ultrasound attenuation in our phantom was 5% higher, respectively, than that specified by QIBA[17], which could have overestimated FibroScan measurements by approximately 7% while other modalities, which encode laterally propagating shear waves, are less affected. The acoustic contrast in our phantom can easily be adapted to that of liver by reducing the concentration of MCC. However, for technical developments idealized imaging parameters are favorable. Therefore, we consider it an advantage that the liquid-liver phantom provides high signal-to-noise ratio, homogenous imaging contrast in both ultrasound and MRI, and MRI relaxation times which are longer than in human liver[20].

In summary, we introduced a comprehensively characterized liquid-liver phantom tailored to the stiffness dispersion of healthy human liver. The phantom is well suited for elastography in ultrasound and MRI and, thus, fosters standardization of liver-stiffness measurements between different systems. The simple production process ensures reproducible properties which do not change considerably with temperature and which remain stable after an initial curing period of seven days. Being a liquid, the phantom has several advantages, including easy shaping to irregular geometries as encountered *in vivo*, and it can be adapted with inclusions and scatterers to meet specific technical requirements.

**Data and materials accessibility statement**

The liquid-liver phantom and reference SWS dispersion data can be made available upon reasonable request to the corresponding author.

**Acknowledgments**

We thank Sanam Assili for preliminary phantom material development and Sophie Schmelzer for performing measurements with Echosense. Funding from the Deutsche Forschungsgemeinschaft (DFG, German Research Foundation, SFB 1340 and BIOQIC GRK2260) is gratefully acknowledged.

**Supplementary material**

**Abbreviations**

QIBA – Quantitative Imaging Biomarkers Alliance

SWS – Shear-wave speed

PAAm – Polyacrylamide

PR – Penetration rate

**Imaging properties**

*Methods*

For the characterization of ultrasound imaging properties, attenuation was measured 10 times per phantom using attenuation imaging of Aplio i900 (Canon, Tokio, Japan) excluding a rim of 2 cm thickness around the phantom boundaries. Speed of sound was measured with SonixMD by inserting the transducer into the phantoms at various depths and comparing the true transducer-bottom distance with the distance displayed in the B-mode. The speed of sound was then obtained from linear regression analysis. Speckle brightness was visually assessed by comparing the CIRS phantom (Elasticity QA phantom model 049, CIRS, Norfolk, USA) with PAAm using the same SonixMD imaging preset (supplementary Figure 1). MRI relaxation times were measured by T1 and T2 mapping sequences using a 3-Tesla MRI scanner (Siemens Magnetom Lumina, Erlangen, Germany). The average value in a manual drawn region of interest was calculated. Image quality was assessed by a T2-weighted sequence.

*Results*

Supplementary Figure 1 reveals that PAAm is more homogenous in MRI than the CIRS phantom. In addition, speckle brightness in ultrasound is higher in PAAm compared to the CIRS phantom, which has similar speckle brightness compared to human liver. However, as discussed in the main text, speckle concentration can be easily modified to match human liver and thereby better fulfil Quantitative Imaging Biomarkers Alliance (QIBA) requirements[17].

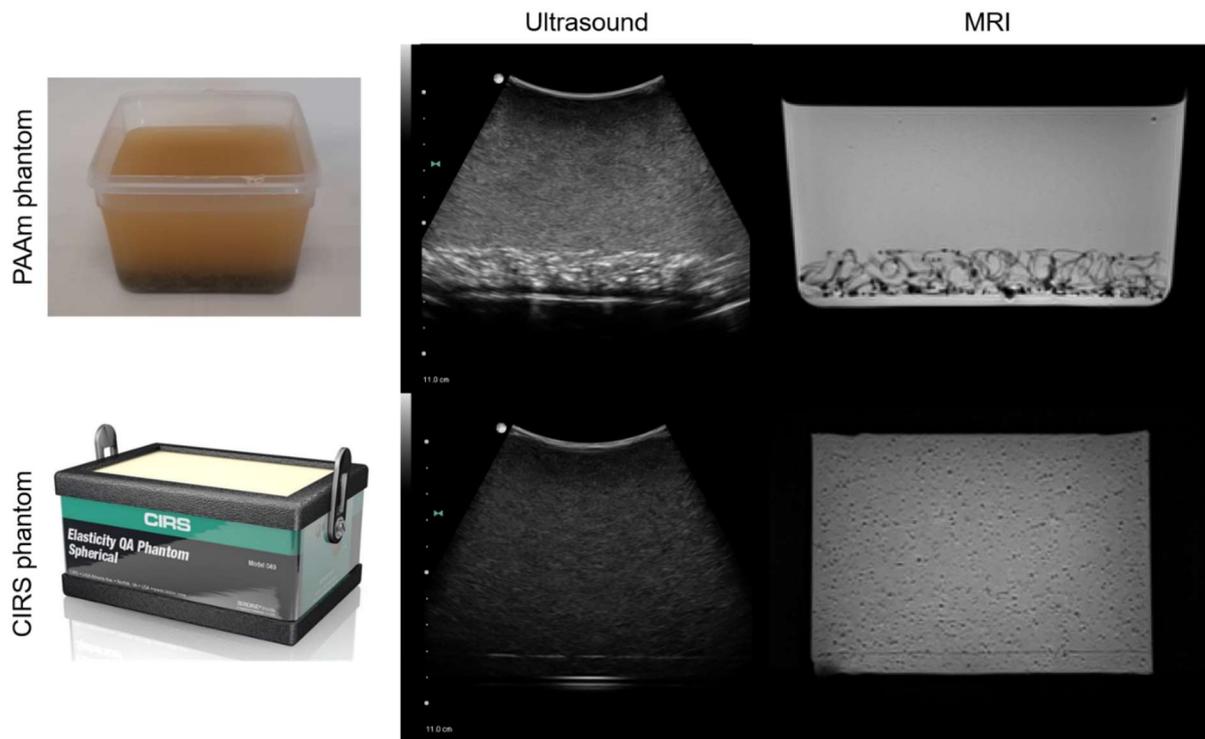

*Supplementary Figure 1* Comparison of the imaging properties of the polyacrylamide (PAAm) phantom and the CIRS phantom with ultrasound and magnetic-resonance imaging (MRI).

**Redistribution of settled scatters**

Supplementary figure 2 shows the uniformly redistribution of segregated microcrystalline cellulose (MCC) scatters.

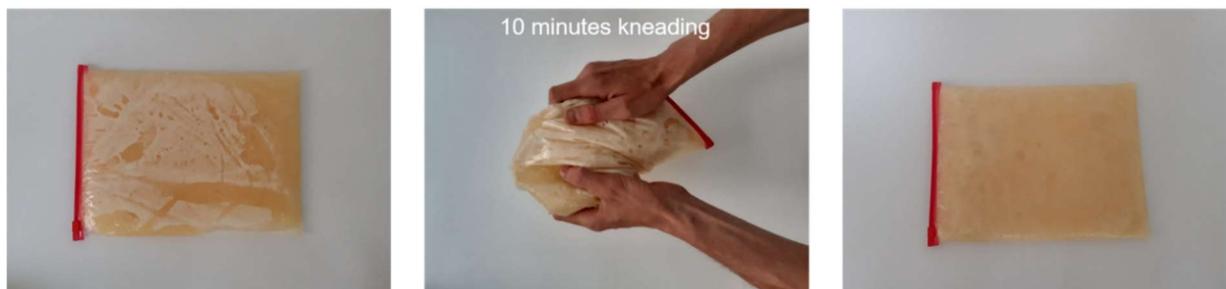

*Supplementary Figure 2* Demonstration of scatter distribution by kneading the phantom bag. Left shows the settled scatters, after 10 minutes kneading (middle), the scatters are redistribute uniformly (right).

**Shear-wave damping**

*Introduction*

To fully characterize the viscoelastic properties of the PAAm phantom, SWS and shear-wave damping were quantified within the frequency range from 5 Hz to 3000 Hz. Therefore, we here briefly introduce penetration rate (PR)[14], which is defined as penetration length (length corresponding to a shear-wave amplitude decay to 1 over the Euler constant) times vibration frequency. PR is inversely related to shear-wave damping. Literature data of PR of the human liver are sparse. To our knowledge, only one study reported PR values of the human liver *in vivo*[21].

*Methods*

All reference elastography methods provided SWS and PR values. Magnitude $|G^*|$ and phase angle of the complex shear modulus $\phi = \arg(G^*)$ obtained from shear rheometry were converted to PR by

$$PR = \frac{1}{2\pi} \sqrt{\frac{2\,|G^*|}{\rho\,[1-\cos(\phi)]}}, \qquad (S1)$$

with PAAm density $\rho$ = (1.094 ± 0.006) g/cm$^3$. To analyze the rheological behavior only for SWS a two-parameter springpot model is sufficient. However, when analyzing SWS and PR dispersion together, the extension to the three-parameter fractional Maxwell model is necessary (assuming a viscosity with $\eta$ = 1 Pa·s),

$$G^*_{\text{FM}} = \frac{\mu_{\text{FM}}^{1-\alpha_{\text{FM}}}\,(i\,\omega\,\eta)^{\alpha_{\text{FM}}} \cdot i\,\omega\,\eta_{\text{FM}}}{\mu_{\text{FM}}^{1-\alpha_{\text{FM}}}\,(i\,\omega\,\eta)^{\alpha_{\text{FM}}} + i\,\omega\,\eta_{\text{FM}}}. \qquad (S2)$$

PR-based characterizations of reproducibility, aging, temperature effect and effect of MCC in PAAm were identical to those based on SWS, which are explained in the main document.

*Results*

Mean PR of all PAAm phantoms were acquired by rheometry, THE, MRE and tabletop MRE. The fractional Maxwell model fit, according to equation S2, yields the parameters $\mu_{\text{FM}}$ = (6.870 ± 0.028) kPa, $\alpha_{\text{FM}}$ = 0.3617 ± 0.0022, and $\eta_{\text{FM}}$ = (60 ± 7) Pa·s. which is shown in supplementary Figure 3. The fit is redrawn in supplementary Figure 4 superposed over the literature data for healthy human liver and the commercial Resoundant phantom (MR Elastography Phantom, Rochester, USA). At reference frequency of 30 Hz, the fractional Maxwell model predicts a PR of (0.583 ± 0.012) m/s. PR reproducibility at 30 Hz in three different phantoms was (4.71 ± 0.06) %. Within the first seven days after phantom production, PR$_{30\,\text{Hz}}$ increased significantly by (15.8 ± 0.7) %, while between days 7 and 17, PR$_{30\,\text{Hz}}$ changed only insignificantly by (5.4 ± 0.5) % (supplementary Figure 5). In the temperature range (26 – 37) °C, PR$_{30\,\text{Hz}}$ decreased significantly by (2.006 ± 0.012) %/°C. Adding MCC significantly increased PR$_{30\,\text{Hz}}$ by (1.80 ± 0.12) %. All parameters are summarized in supplementary Table 1.

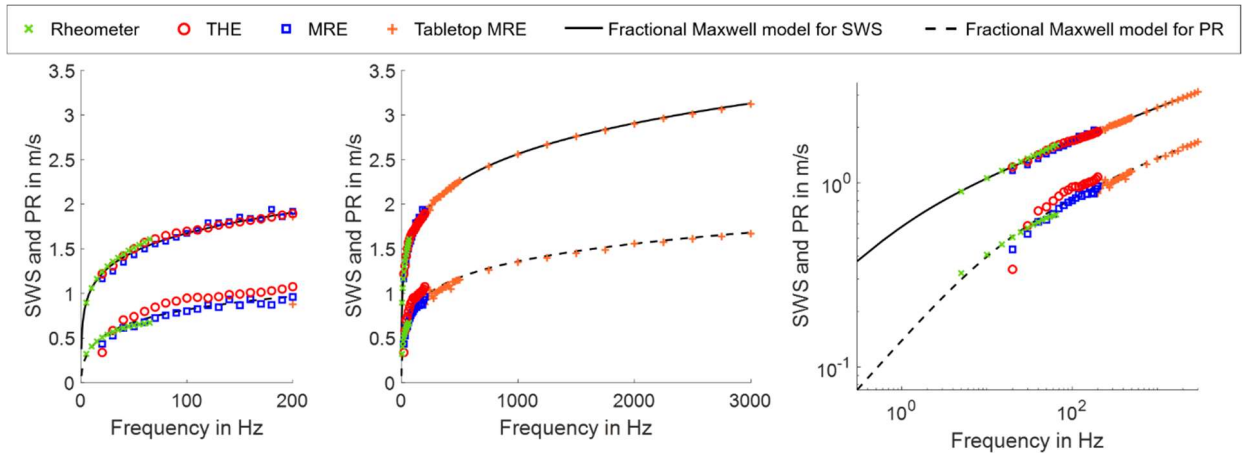

*Supplementary Figure 3* Multimodal characterization of shear-wave speed (SWS) and penetration rate (PR) dispersion of the PAAm phantom. Data are fitted by the fractional Maxwell model for SWS (black solid line) and for PR (black dashed line). Left-hand side: frequency zoom 0 Hz to 200 Hz; middle: full frequency range, normal scale; right-hand side: full frequency range, log scale. Please note that each modality provides both SWS and PR in m/s.

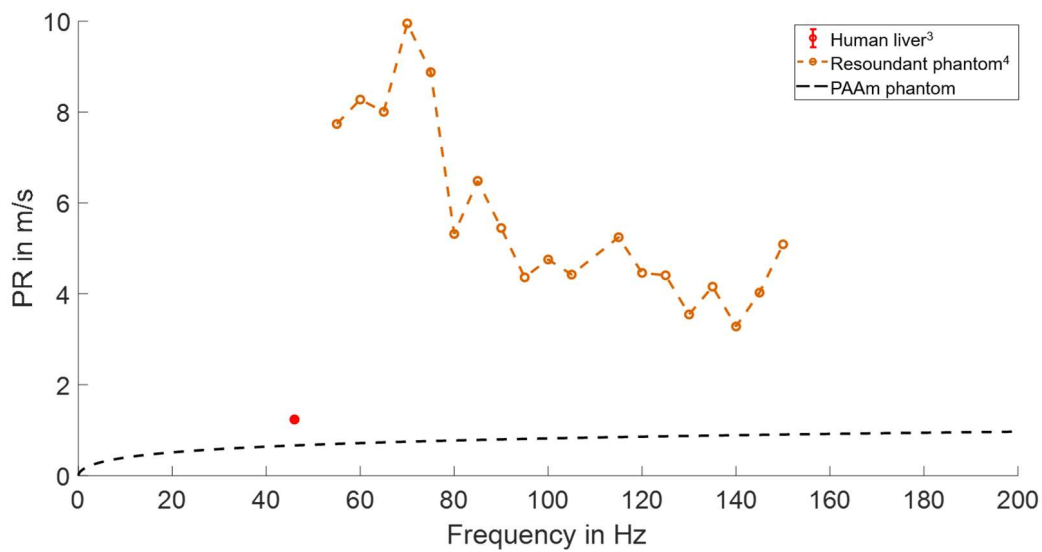

*Supplementary Figure 4* Overview of frequency-resolved penetration rate (PR), literature data on healthy human liver *in vivo*[3] and the commercial MRE phantom (Resoundant)[4]. In addition, the fitted PR dispersion of the proposed PAAm phantom is redrawn from supplemental figure 3. Magnetic-resonance elastography (MRE)

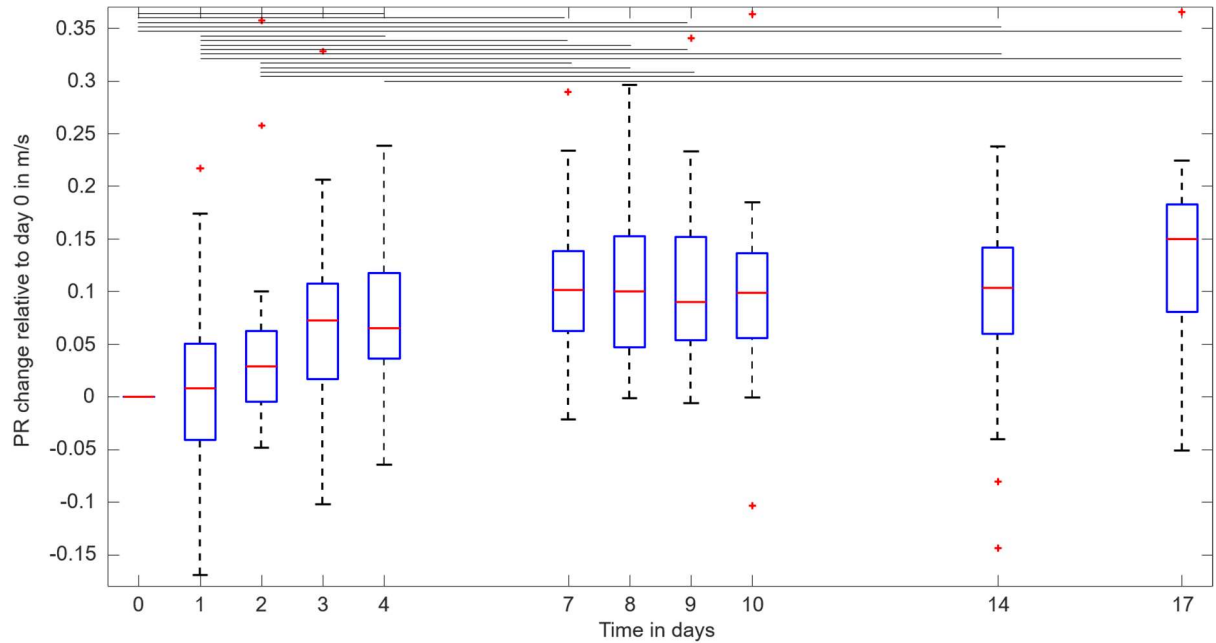

*Supplementary Figure 5* Analysis of penetration rate (PR) changes due to aging based on tabletop MRE in the frequency range from 200 Hz to 3000 Hz. PR changes relative to production (day 0) are shown. The variability of data due to frequencies is indicated by the boxplots. Horizontal black lines demarcate significant changes.

| Group | Type | Value | | | | P-value | PR at 30 Hz | Requirements |
|---|---|---|---|---|---|---|---|---|
| | | Parameter | Unit | Mean | SE | | | |
| Viscoelastic properties | Rheological model | $\alpha_{FM}$ | | 0.3617 | 0.0022 | | 0.583 ± 0.012 m/s | |
| | | $\mu_{FM}$ | kPa | 6.870 | 0.028 | | | |
| | | $\eta_{FM}$ | Pa s | 60 | 7 | | | |
| | | | | | | | **Relative effect at 30 Hz** | |
| | Reproducibility | PR range | m/s | 0.0549 | 0.0007 | | ± 4.71 ± 0.06 % | ± 5 % |
| | Aging effect | $PR_{Day\,7} - PR_{Day\,0}$ | m/s | 0.092 | 0.004 | 0.006 | 15.8 ± 0.7 % / 7 days | < 5 % / 6 month |
| | | $PR_{Day\,17} - PR_{Day\,7}$ | m/s | 0.0312 | 0.0028 | 0.12 | 5.4 ± 0.5 % / 10 days | |
| | Temperature effect | $PR_{37\,°C} - PR_{26\,°C}$ | m/s/°C | -0.01170 | 0.00007 | < 0.0001 | - 2.006 ± 0.012 % / °C | 21 ± 1 °C |
| | MCC effect | $PR_{MCC} - PR_{No\,MCC}$ | m/s | 0.0105 | 0.0007 | 0.01 | 1.80 ± 0.12 % | ± 5 % |

*Supplementary Table 1* Viscoelastic properties of the polyacrylamide (PAAm) phantom characterized by penetration rate (PR) with specifications from the Quantitative Imaging Biomarkers Alliance[1]. Standard error (SE)

## Discussion

We analyzed shear-wave damping in terms of PR (please note the inverse relationship between PR and damping). Shear-wave damping of our PAAm phantom was slightly higher compared with healthy human liver *in vivo*. However, this higher damping better represents the scenario of liver tissue than that mimicked by Resoundant or CIRS phantoms which have much lower or even zero shear-wave damping[4]. For transient shear-wave excitation, the higher damping affects the estimated group frequency by a shift towards lower frequencies.

Crosslinking and mixing crosslinked and non-crosslinked (linear polarized) PAAm chains decreases shear-wave damping compared to non-crosslinked PAAm as demonstrated by preliminary data (Supplementary Figure 6). The crosslinking was performed according to Charrier et al., 2018[22]. It is visible that crosslinking changes the PAAm from a more liquid to a more solid state which better reproduces liver SWS in the limit of static load by yielding an SWS offset. Also, PR at that limit is infinitely high for the crosslinked PAAm as it is predicted for a solid while it is zero for the non-crosslinked PAAm as expected for a liquid while the fully crosslinked PAA cannot mimic liver viscoelastic dispersion, mixing non-crosslinked with crosslinked PAAm offers a way to balance between static solid properties and viscoelastic dispersion. We noted that our preliminary mixtures of non-crosslinked and crosslinked PAAm could mimic the realistic SWS and PR dispersion of *in vivo* elastography as shown in supplementary figure 6 and provide a realistic static shear modulus at the same time. Further research is needed to complement our proposed liquid liver phantom with a solid phantom that also meets the criteria for liver elastography, including reproducibility and the required imaging properties.

Future studies of the frequency resolved damping properties of the human liver *in vivo* are necessary to provide further reference data. Since QIBA requirements[17] are stated explicitly only for SWS, the measured PR properties can only be compared with SWS properties. Here, all effects addressed in our study, expect the effect of aging, comply with the requirements. Although our phantom was specifically tailored to SWS dispersion observed in *in vivo* liver tissue, all PR values are in better agreement with liver tissue than those of other phantoms.

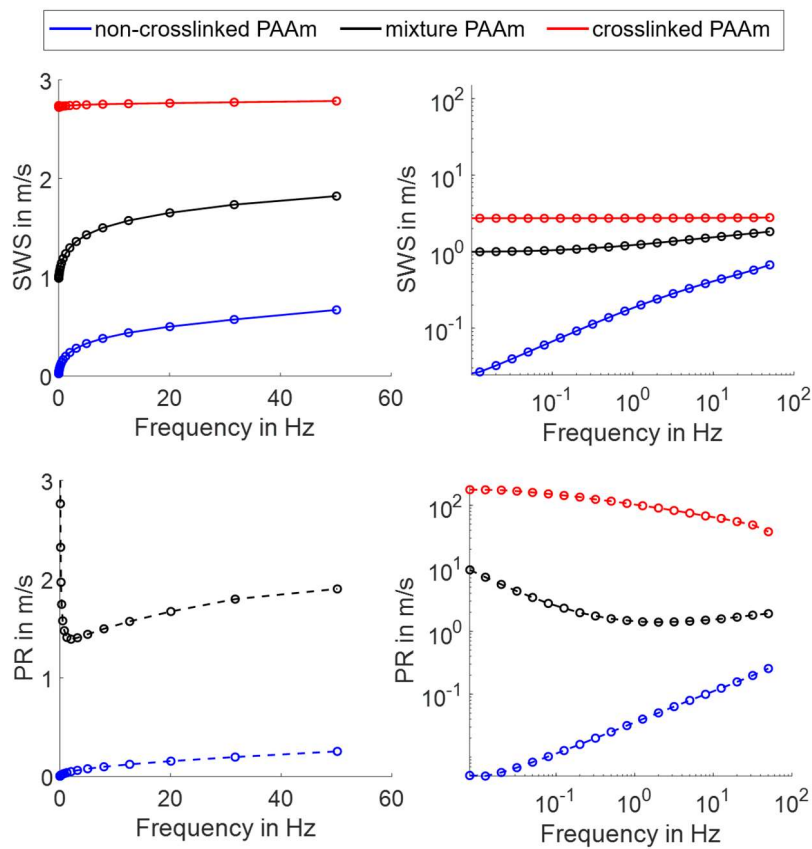

*Supplementary Figure 6* Shear-wave speed (SWS, on the top) and penetration rate (PR, on the bottom) shown for a non-crosslinked polyacrylamide (PAAm, blue) phantom and a mixture PAAm phantom (black) in comparison with a crosslinked PAAm (red). On the left full frequency range on the right log scale.

**References for supplementary material**